\documentclass[a4paper,11pt]{article}
\usepackage[margin=3cm]{geometry}

\usepackage{microtype}
\usepackage{amssymb,amsmath,amsopn,amssymb,amsfonts,amsthm}
\usepackage{authblk}
\usepackage{cite}
\usepackage{pifont}
\usepackage{complexity}
\usepackage{paralist}

\theoremstyle{plain}
\newtheorem{theorem}{Theorem} 

\newtheorem{obs}{Observation}

\usepackage{graphicx}
\graphicspath{{figures/}}

\usepackage{xspace}
\usepackage{anyfontsize}
\usepackage[inline]{enumitem}
\usepackage{subcaption}

\usepackage{xcolor}
\definecolor{darkblue}{rgb}{0,0,0.4}
\definecolor{darkred}{rgb}{0.5,0,0}

\usepackage{hyperref}
\hypersetup{colorlinks=true,
    linkcolor=darkblue,
    citecolor=darkred,
    urlcolor=blue}

\newcommand{\true}{\textsc{true}\xspace}
\newcommand{\false}{\textsc{false}\xspace}
\renewcommand{\SAT}{\textsc{SAT}\xspace}

\makeatletter
\renewcommand\paragraph{\@startsection{paragraph}{4}{\z@}%
                                      {\parskip}
                                      {-1em}%
                                      {\normalfont\normalsize\bfseries}}
\makeatother

\title{The Partition Spanning Forest Problem\thanks{This work started at the 14th European Research Week on Geometric Graph (GGWeek'17) in Vierhouten, The Netherlands. A preliminary version was presented at the 34th European Workshop on Computational Geometry (EuroCG'18)~\cite{kkrss-tpsfp-eurocg18}. This research was funded in part by Humility \& Conviction in Public Life, a project of the University Connecticut sponsored by the John Templeton Foundation.}}

\author[1]{Philipp~Kindermann}
\author[2]{Boris~Klemz}
\author[3]{Ignaz~Rutter}
\author[4]{Patrick~Schnider}
\author[5]{Andr\'e~Schulz}
\affil[1]{
  University of Waterloo, Canada, 
  \texttt{philipp.kindermann@uwaterloo.ca}}
\affil[2]{
  Freie Universit\"at Berlin, Germany,
  \texttt{klemz@inf.fu-berlin.de}}
\affil[3]{
  University of Passau, Germany,
  \texttt{rutter@fim.uni-passau.de}}
\affil[4]{
  ETH Z\"urich, Switzerland,
  \texttt{patrick.schnider@inf.ethz.ch}}
\affil[5]{
  FernUniversit\"at in Hagen, Germany,
  \texttt{andre.schulz@fernuni-hagen.de}}

\begin{document}
\maketitle


\begin{abstract}
Given a set of colored points in the plane, we ask if there 
exists a crossing-free straight-line drawing of a spanning forest, 
such that every tree in the forest contains exactly the points of
one color class. We show that the problem is
$\NP$-complete, even if every color class contains at most five
points, but it is solvable in $O(n^2)$
time when each color class contains at most three points.
If we require that the spanning forest is a linear forest,
then the problem becomes $\NP$-complete even if every color class contains at most four points.
\end{abstract}

\section{Introduction}\label{sec:introduction}

\begin{figure}[t]
  \centering
  \includegraphics{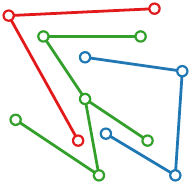}
  \caption{A solution to a problem instance with three color classes.}
  \label{fig:example}
\end{figure}

Let $P=\{p_1,\dots,p_n\}$ be a set of~$n$ points in the plane
and let $C=\{C_1,\ldots,C_k\}$ be a partition of~$P$ into~$k$
sets of points, called \emph{color classes}, such that every point 
belongs to exactly one color class. We study the \emph{partition
spanning forest problem} which is defined as follows:
Is there a crossing-free straight-line drawing of a spanning forest~$F$
that consists of~$k$ trees~$T_1,\ldots,T_k$ such that each tree~$T_i$, $1\le i\le k$,
contains exactly the points of the color class~$C_i$?
Figure~\ref{fig:example} shows an example with three color classes.

For $k=1$, the problem is equivalent to finding a geometric spanning
tree of~$P$ which trivially always exists. Hence, several optimization
versions of this problem have been studied in the past; see Eppstein~\cite{e-sts-hcg99}
for a survey. Bereg et al.~\cite{bjyz-rbstp-TCS11} showed how to solve the 
problem in $O(n\log n)$ time in the case of $k=2$. Hiu and
Schaefer~\cite{hs-ppt-isaac04} proved that it is $\NP$-complete to decide for
two color classes $A=\{a_1,\ldots,a_n\}$ and $B=\{b_1,\ldots,b_n\}$ whether
there exists an ordering~$\pi$ such that the geometric paths $a_{\pi_1},\ldots,a_{\pi_n}$
and $b_{\pi_1},\ldots,b_{\pi_n}$ are crossing-free. 
Bereg et al.~\cite{bfkpsw-cnesf-isaac15} asked for not necessarily straight-line
Steiner trees for each color class of minimum total length and gave a PTAS for $k=2$,
a $(5/3+\varepsilon)$-approximation for $k=3$, 
and a $(k+\varepsilon)$-approximation for $k>2$.

In this paper, we analyze the complexity of the partition spanning
forest problem for color classes of bounded size. We give an
$O(n^2)$-time algorithm when each color class contains at most three
points (Sec.~\ref{sec:3pts}) and show that the problem is
$\NP$-complete for up to five points per color class
(Sec.~\ref{sec:5pts}); the complexity for four points remains open.
In Section~\ref{sec:linear}, we show that the \emph{partition spanning
  linear forest problem}, where each tree is required to be a path, is
$\NP$-complete, even if every color class contains at most four points.  
The complexity of the non-linear version remains open if
every color class contains at most four points.

\section{Color classes with at most three points}\label{sec:3pts}
In the case where each color class of the input instance
contains of at most three points, the partition spanning forest
problem can be solved in polynomial time. In fact,
with this restriction the problem can be formulated as a 
2-\SAT problem.

Assume that our point set $P = \{p_1,\dots,p_n\}$ consists of $n$
points.  In the following we will 
understand the color classes as subsets
$I \subseteq [n] := \{1,\dots,n\}$ of indices.  For a point $p_i$ we
denote its color class by $I(p_i)$.  We refer to the edges
$(p_i,p_j)$ where $p_i$ and $p_j$ are in the same color class as the
\emph{potential edges} of the instance.  Observe that an arbitrary
choice of the potential edges forms a solution to the problem 
(with at most three points per color class) if and
only if it satisfies the following conditions:
\begin{inparaenum}[(i)]
\item For each point $p_i$, if $|I(p_i)| > 1$, then at least one
  potential edge incident to~$p_i$ must be chosen.
\item For any pair of potential edges $p_ip_j$ and $p_kp_l$ that
  intersect in the interior, at most one of them is chosen.
\item For any color class $I$ with $|I| = 3$ one of the
  potential edges of that color is not chosen.
\end{inparaenum}

Observe that condition (iii) can be skipped, as any choice of
potential edges satisfying conditions (i) and (ii) can be 
extended to also satisfy (iii).

We model the possible choices of potential edges that satisfy
conditions (i) and (ii) by a 2-\SAT formula as follows.  For each
potential edge $(p_i,p_j)$ there is a variable $x_{ij}$ with the
interpretation that if $x_{ij}$ is true, then the edge connecting
$p_i$ to $p_j$ is \emph{not} chosen as part of the solution, and otherwise it
is.

Conditions (i) and (ii) can be expressed as 2-\SAT formulas using the
variables $x_{ij}$ as follows.  For condition (i), we create for each
point $p_i$ the (sub)formula
$\bigvee_{j \in I(p_i) \setminus \{i\}} \lnot x_{ij}$.  Note that
this is a 2-\SAT formula since $|I(p_i) \setminus \{i\}| \le 2$ by the
assumption that each color class has size at most three.  For any two
potential edges $(p_i,p_j)$ and $(p_k,p_l)$ that cross, we add the
clause $x_{ij} \vee x_{kl}$, thus enforcing condition (ii).  It
follows that the resulting 2-\SAT formula $\varphi$ is satisfiable if
and only if the original instance of the partition spanning forest
problem admits a solution.

If a color class contains only one point, then we can always draw 
it as a singleton point since we assumed general position
for our input points. Thus, we are left with sets of either
two or three points. In the case of two points, there is a
unique spanning tree. However, for sets with three points
we have three choices. For each of those spanning trees, we
introduce a boolean variable. In particular, if the color
class is $\{p_i,p_j,p_k\}$, then we denote the boolean variable
for the spanning tree formed by the edges $(p_i,p_j)$ 
and $(p_j,p_k)$ by $x_{ik}$, using its endpoints as the
indices. The interpretation of the variable assignment
will be the following: if $x_{ik}$ is true, then the corresponding 
spanning tree is selected as the spanning tree for its color class; if $x_{ik}$ is false,
then any of the three possible spanning trees of its
color class can be chosen. To make this work, we have to 
guarantee that at most one of $x_{ij},x_{jk}$, and $x_{ik}$
is true. This can be enforced by the 2-\SAT
(sub)formula 
\[(\lnot x_{ij} \lor \lnot x_{ik}) \land
(\lnot x_{jk} \lor \lnot x_{ik}) \land
(\lnot x_{ij} \lor \lnot x_{jk}).\]
We add this formula for every color class with three elements.

In the next step, we process each pair of color classes.
While processing, we will observe one of the following:
(i) the local configuration already forbids the existence of a 
partition drawing, (ii) the two sets impose a constraint on
the available spanning trees, or 
(iii) the two sets do not interfere with each other.
In case of (i) we can stop the algorithm, in case of (ii) 
we (iteratively) build a 2-\SAT formula to model these constraints.


Let now $A$ and $B$ be a pair of color classes. 
If $|A|=|B|=2$, then their convex hull is either intersecting or not. 
In the former case, there exists no partition drawing; in the
latter, these two sets impose no constraints.

If one of the color classes contains three points 
(say~$A$) and the other contains two points, then we 
are left with one of the following situations. The convex hulls
of both sets could be disjoint, which yields no constraints.
If two edges of the convex hull of~$A$ are intersected by the convex
hull of~$B$, then there cannot be a spanning tree of~$A$ avoiding
the edge spanned by~$B$. Thus, in this case we cannot have
a partition drawing. Finally, if the segment spanned by~$B$ 
intersects a single edge of the convex hull of~$A$, then
only one of the three possible spanning tress of~$A$
can be part of a partition drawing. In this case, we add an
appropriate clause to the 2-\SAT formula that enforces
the corresponding spanning tree.

\begin{figure}[t]
  \centering
  \subcaptionbox{\label{fig:2sat-1}}{\includegraphics[page=1]{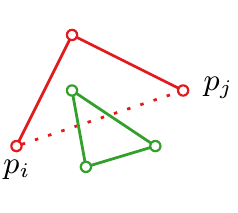}}
  \hfil
  \subcaptionbox{\label{fig:2sat-2}}{\includegraphics[page=2]{2sat}}
  \hfil
  \subcaptionbox{\label{fig:2sat-3}}{\includegraphics[page=3]{2sat}}
  \caption{Local constraints for spanning trees. (a)~A situation
  where one color class is restricted to one spanning tree. (b--c)~Another situation where two depended spanning trees for each color class are possible.}
  \label{fig:2sat}
\end{figure}

We are left with the case that $|A|=|B|=3$.  Clearly, if the convex
hulls of these sets are disjoint, then this pair imposes no constraints.
If their convex hulls intersect in four or even six points, it is an
easy exercise to see that in this case a partition drawing is not
possible. If there are two intersection points, we have 
to consider two cases.
If both intersections lie on the same edge, say $(p_i,p_j)$ spanned by points
from~$A$, then only one spanning tree in~$A$ can be chosen (see Figure~\ref{fig:2sat}(a)). In this
case, we enforce $x_{ij}$ to be true by adding the clause $x_{ij}$
to the formula. In the remaining case, let $A=\{p_i,p_j,p_k\}$ and $B=\{p_a,p_b,p_c\}$. 
We assume that $(p_i,p_j)$ intersects
$(p_a,p_b)$ and that $(p_j,p_k)$ intersects $(p_b,p_c)$. Now,
we have two pairs of possible spanning trees (see Figure~\ref{fig:2sat}(b--c)). To model
this, we add the clauses
\[ (\lnot x_{ij} \lor \lnot x_{ab}) \land (\lnot x_{jk} \lor \lnot x_{bc})\]
to our 2-\SAT formula.

By the above strategy, we have constructed a 2-\SAT 
formula that is satisfiable if and only if the input instance 
has a partition drawing. 

The formula has length at most $O(n^2)$
and can be constructed in $O(n^2)$ time as well. 
By using an efficient algorithm for 2-\SAT~\cite{APT79}, we get the
desired algorithm. We summarize our construction in the following
theorem.

\begin{theorem}
	The partition spanning forest
problem for $n$ points can be solved in $O(n^2)$ time if every 
color class contains at most three points.
\end{theorem}

\section{Color classes with at most five points}\label{sec:5pts}

In this section we prove the following theorem:

\begin{theorem}\label{thm:5pts}
	The partition spanning forest
problem is $\NP$-complete, even if every color class contains at most five points.
\end{theorem}

The problem is obviously contained in~$\NP$. In order to show
the~$\NP$-hardness, we perform a polynomial-time reduction from
\textsc{Planar 3-Satisfiability}.  In this $\NP$-hard~\cite{l-pftu-82}
special case of 3SAT the input is a 3SAT formula~$\varphi$ whose
variable--clause graph is planar.  We can assume that such a formula is
given together with a contact representation~$\mathcal R$
of~$\varphi$~\cite{kr-pcr-92}.  Thus, all variables are represented as
horizontal line segments arranged on one line. Each clause~$c$ is
represented as an E-shape turned by $90^\circ$ such that the three
vertical \emph{legs} of the E-shape touch precisely the variables
contained in~$c$.
For our reduction, we construct a set of colored points that admits a partition drawing if and only if~$\varphi$ is satisfiable.

\paragraph{Overview.}
We introduce five types of gadgets.
For each variable~$u$ we create a \emph{variable gadget} which admits exactly two distinct partition drawings.
These drawings correspond to the two truth states of~$u$.
\emph{Wire gadgets} are used to propagate these states to the \emph{clause gadgets}, one of which is created for every clause~$c$.
The clause gadget of~$c$ ensures that gadget configurations of the variables contained in~$c$ correspond to a truth assignment in which at least one of the literals of~$c$ is satisfied.
In order to connect our gadgets appropriately we also require a \emph{splitting gadget}, which splits one wire into two wires, and we require a gadget that flips the state transported along a wire.
We proceed by describing our gadgets in detail.
Note that different gadgets always use different color classes, even if we might give them the same name in the construction (so there are many \emph{red} color classes in an instance).

\begin{figure}[t]
  \centering
  \subcaptionbox{\label{fig:5pts-hardness-false}State (i)}{\includegraphics[page=4]{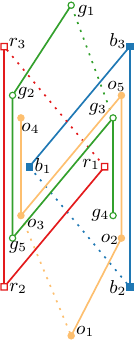}}
  \hfil
  \subcaptionbox{\label{fig:5pts-hardness-true}State (ii)}{\includegraphics[page=5]{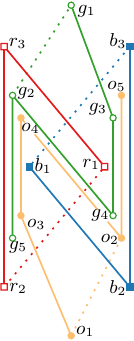}}
  \hfil
  \subcaptionbox{\label{fig:5pts-propagation}A chain}{~\includegraphics[page=7]{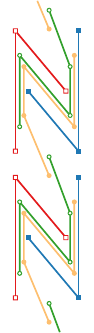}~}
  \caption{The configurations of the wire gadget}
  \label{fig:5pts-hardness}
\end{figure}

\paragraph{The wire gadget.}
The wire gadget consists of four color classes; see Figure~\ref{fig:5pts-hardness}.
The points of the \emph{red} color class $R=\lbrace r_1,r_2,r_3\rbrace$ and the \emph{blue} color class $B=\lbrace b_1,b_2,b_3\rbrace$ are arranged such that the convex hulls of~$R$ and~$B$ intersect in the two points $b_1b_2\cap r_1r_2$ and $b_1b_3\cap r_1r_3$.
As a consequence, there are exactly two possible configurations for the red and blue spannings trees which can be used in a partition drawing, see Figure~\ref{fig:5pts-hardness-false} and Figure~\ref{fig:5pts-hardness-true}.
Either choice uniquely determines the spanning tree of both the \emph{green} color class $G=(g_1,\dots,g_5)$ and the \emph{orange} color class $O=(o_1,\dots,o_5)$, as the edges of the red and blue spanning trees obstruct all other possible green and orange edges.
Thus, there are exactly two possible partition drawings of the wire gadget.
In particular, these two drawings satisfy the following.

\begin{obs}\label{thm:5pt-wire}
Any partition drawing of the wire gadget either contains (i) the edges $g_1g_2$ 
and $o_1o_2$, but not the edges $g_1g_3$ and $o_1o_3$, see 
Figure~\ref{fig:5pts-hardness-false};  or (ii) the edges $g_1g_3$ and $o_1o_3$, but not the 
edges $g_1g_2$ and $o_1o_2$, see Figure~\ref{fig:5pts-hardness-true}.
\end{obs}

These two states (i) and (ii) may be propagated by creating \emph{chains} of
wire gadgets in which the convex hulls of consecutive gadgets intersect in two 
points as illustrated in Figure~\ref{fig:5pts-propagation}.
Consider two consecutive wire gadgets in a chain.
By Observation~\ref{thm:5pt-wire}, either both gadgets are in state (i) or both gadgets are in state (ii) due to the way their convex hulls intersect.
As a consequence, the first gadget of the chain is in state (i) if and only if the last one is in state (i) as well.
Chains are flexible structures and turns can easily be implemented by curving a chain.
Further, the length of a chain may be adjusted by increasing or decreasing the distance between consecutive wire gadgets.

\begin{figure}[t]
  \centering
  \includegraphics[trim=1cm 3cm 1cm 3cm,clip,page=2]{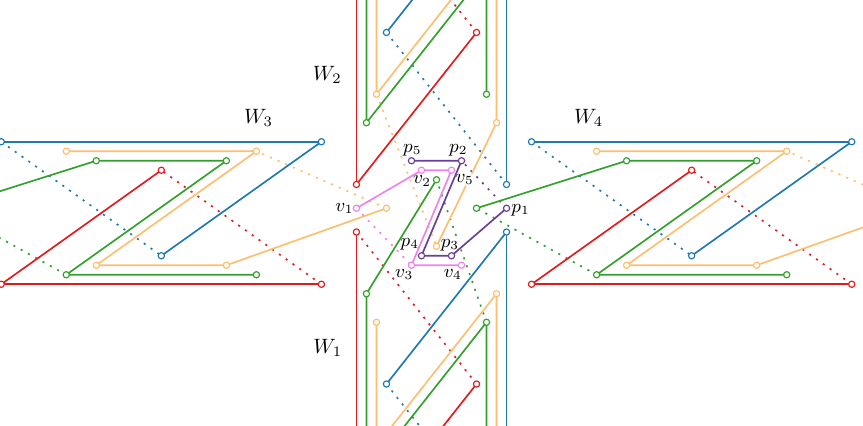}
  \caption{Splitting gadget}
  \label{fig:5pt-splitter}
\end{figure}

\paragraph{Splitting and inverting.}
The splitting gadget consists of two color classes $V=\lbrace v_1,\dots,v_5\rbrace$ (\emph{violet}) and $P=\lbrace p_1,\dots,p_5\rbrace$ (\emph{purple}) whose points are placed between two consecutive wires $W_1,W_2$ in a chain, see Figure~\ref{fig:5pt-splitter}.
The functionality of these two color classes is similar to the one of the color classes green and orange in the wire gadget:
the state of~$W_1$ and~$W_2$ uniquely determines the spanning tree of both the violet and the purple color class.
In particular, the purple tree contains either~$p_1p_3$ or~$p_1p_5$ and the violet tree contains either~$v_1v_2$ or~$v_1v_3$.
We may now attach one or two additional wires perpendicular to the chain such that their convex hulls intersect the convex hull of the splitting gadget, see $W_3$ and $W_4$ in Figure~\ref{fig:5pt-splitter}.
The edges incident to~$p_1$ and~$v_1$ in the purple and violet spanning trees allow precisely one state for both~$W_3$ and~$W_4$.

\begin{obs}
\label{thm:5pt-split}
In any drawing of the splitting gadget, the state of the wires~$W_3$ and $W_4$ differs from the state of~$W_1$ and~$W_2$.
\end{obs}

In this sense, the splitting gadget does not only split a wire into two wires, it can also be used to flip the state propagated along a chain.

\paragraph{The variable gadget.}
The variable gadget is a horizontal chain to which we attach multiple wires using splitters.
The number of wires attached from the top (bottom) matches the number of E-shape legs touching the variable from the top (bottom) in the contact representation~$\mathcal R$ of~$\varphi$.

\begin{figure}[b]
    \centering
    \subcaptionbox{$\ell_1=\ell_2=\ell_3=\true$\label{fig:5pts-clause-gadget-alltrue}}{\includegraphics[page=1]{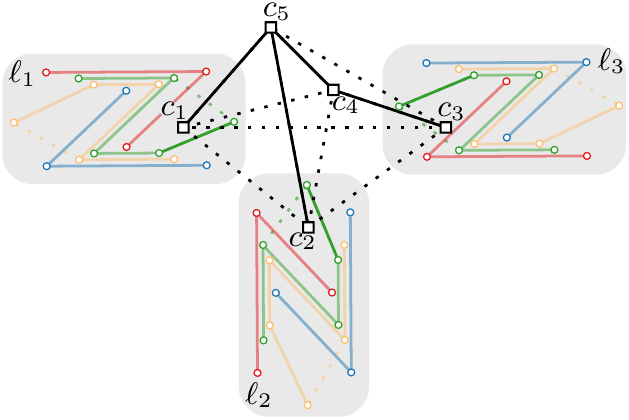}}
    \hfil
    \subcaptionbox{$\ell_1=\ell_2=\ell_3=\false$\label{fig:5pts-clause-gadget-allfalse}}{\includegraphics[page=8]{5pts-clause-gadget}}
    \caption{The clause gadget between literals~$\ell_1,\ell_2,\ell_3$.}
    \label{fig:5pts-clause-gadget}
\end{figure}

\paragraph{The clause gadget.}
The clause gadget for a clause of three literals~$\ell_1,\ell_2,\ell_3$ consists of 
one color class with exactly five vertices~$c_1,\ldots,c_5$. We place~$c_1,c_2$,
and~$c_3$ inside a wire gadget representing~$\ell_1,\ell_2$, and~$\ell_3$,
respectively, and we place~$c_4$ and~$c_5$ between those as depicted in Figure~\ref{fig:5pts-clause-gadget}.
We will now show that the gadget is drawable if and only if at least one of~$\ell_1,\ell_2,\ell_3$
is~\true. In particular, we can always use an edge to connect~$c_4$ and~$c_5$.
We can connect~$c_3$ to~$c_4$ if~$\ell_3$ is~\true and we can connect~$c_3$
to~$c_5$ otherwise; similarly, we can connect~$c_2$ to~$c_5$ if~$\ell_2=\true$ 
and we can connect~$c_2$ to~$c_4$ otherwise. If~$\ell_1=\true$, then we
can always connect~$c_1$ to~$c_5$. However, if~$\ell_1=\false$, then we cannot
connect~$c_1$ to~$c_4$ or~$c_5$, and we can connect it to~$c_2$ or~$c_3$
only if~$\ell_2$ or~$\ell_3$ is~\true, respectively. Hence, the gadget is not
drawable if $\ell_1=\ell_2=\ell_3=$\false. Note that the connection from~$c_1$
to~$c_3$ might intersect the connection from~$c_2$ to~$c_4$. However, we
only have to use it if~$\ell_1=\ell_2=\false$ and~$\ell_3=\true$; in this
case, we can connect~$c_2$ to~$c_3$ instead of~$c_4$. Thus, the gadget
is drawable if and only if at least one of~$\ell_1$, $\ell_2$, and~$\ell_3$ is \true.

\begin{figure}[t]
  \centering
  \includegraphics{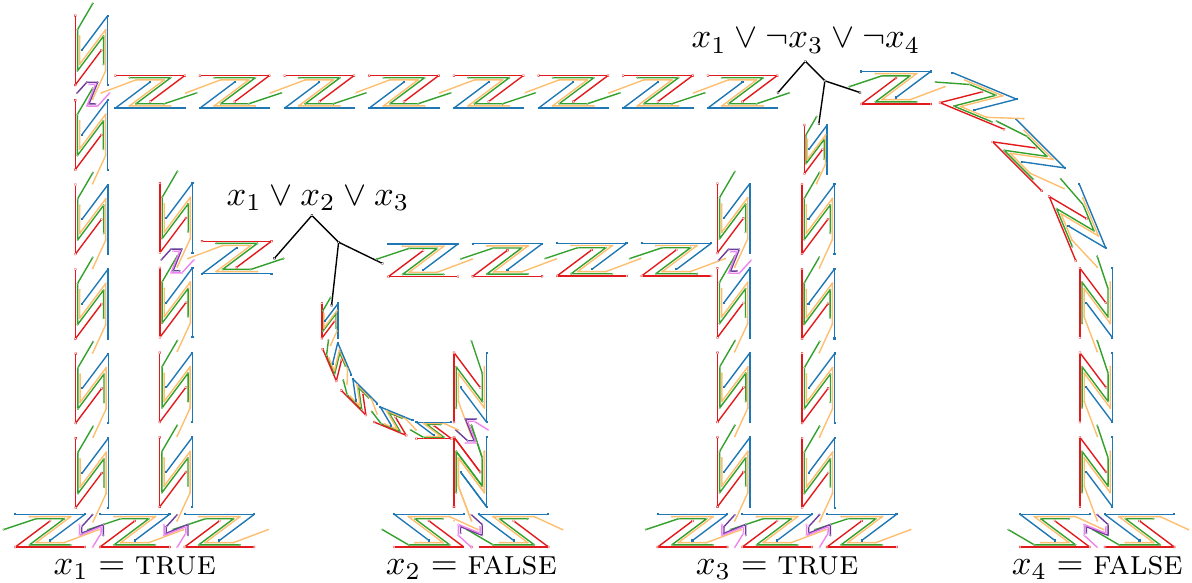}
  \caption{A full example.}
  \label{fig:5pts-example}
\end{figure}

\paragraph{Layout and correctness.}
The wires that are attached to the variable gadgets are vertical and,
by Observation~\ref{thm:5pt-split}, their state is inverted, so they propagate
the negated variable. Hence, if a literal is positive, we have to invert the 
state of the wire again. Two of the wires are supposed to enter the clause 
horizontally; for these two, if they correspond to a positive literal,
we simply use another splitting gadget to make the wire horizontal. Otherwise,
the wire makes a $90^\circ$ degree turn to become horizontal and to
propagate the negated variable. The third wire is supposed to enter the clause gadget
vertically, so if its literal is negative, the vertical wire can directly
connect to the clause. Otherwise, we use another splitting gadget followed
by a $90^\circ$ degree turn. See Figure~\ref{fig:5pts-example} for an example
of that shows all cases. Since the clause gadgets are drawable if and only if
one of their literals is \true and since the wires propagate the states of
the variable gadgets, the resulting instance is drawable if and only if the
planar 3SAT formula~$\varphi$ is satisfiable,
which proves the correctness of Theorem~\ref{thm:5pts}.

\section{Linear forests for color classes with at most four points}\label{sec:linear}

In this section we consider the additional restriction that the spanning forest is a linear forest, that is, each connected component is a path.
Note that, if every color class contains at most three points, then every spanning forest is linear, so in this case we can solve the problem in polynomial time.
On the other hand, we show that under this additional restriction, the problem is $\NP$-complete already if every color class contains at most four points.

\begin{theorem}
	The partition spanning linear forest
problem is $\NP$-complete, even if every color class contains at most four points.
\end{theorem}

Again, the problem is clearly contained in~$\NP$.
In order to show the~$\NP$-hardness, we again perform a polynomial-time reduction from \textsc{Planar 3-Satisfiability}, but using different gadgets.
As before, we construct a variable gadget, a splitting gadget, a wire gadget, and an inverter gadget.
Instead of directly constructing a clause gadget, we will however construct an OR-gadget.
The clause gadget can then be built by concatenating two OR-gadgets and enforcing the resulting variable gadget to be set to \true by crossing the appropriate edge with a new color class consisting of two points.

\paragraph{The variable, wire, and inverter gadgets.}
The variable gadget consists of one color class, the \emph{black} color class $X=\lbrace x_1,x_2,x_3\rbrace$.
Using a second color class, the \emph{blue} color class $B=\lbrace b_1,b_2,b_3\rbrace$, we can enforce that the edge $x_1x_2$ must be drawn in any partition drawing.
The classes $B$ and $X$ are placed in such a way that their convex hulls intersect in two points.
In particular, there are two distinct partition drawings for $B$ and $X$, corresponding to two truth states and $x_1x_2$ is present in both of them.

\begin{figure}[b]
  \centering
  \includegraphics{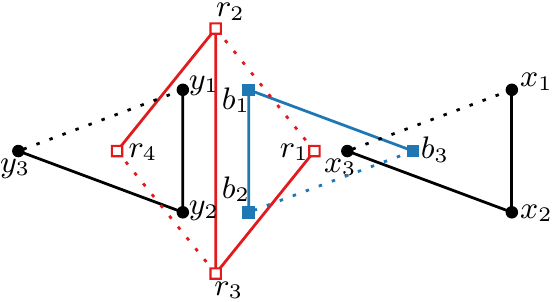}
    \caption{The wire gadget.}
  \label{fig:wire4}
\end{figure}

\begin{figure}[t]
  \centering
  \subcaptionbox{}{
  \includegraphics[page=1]{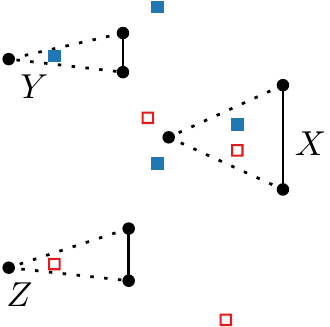}
  }
  \hfil
  \subcaptionbox{}{
  \includegraphics[page=2]{4pts-split}
  }
  \hfil
  \subcaptionbox{}{
  \includegraphics[page=3]{4pts-split}
  }
  \caption{The splitting gadget and its assignments.}
  \label{fig:split4}
\end{figure}

The wire gadget consists of four color classes, the \emph{red} color class $R=\lbrace r_1,r_2,r_3,r_4\rbrace$ and the \emph{blue} color class $B=\lbrace b_1,b_2,b_3\rbrace$, and two \emph{black} color classes $X=\lbrace x_1,x_2,x_3\rbrace$ and $Y=\lbrace y_1,y_2,y_3\rbrace$, see Figure~\ref{fig:wire4}.
Classes $B$ and $X$ are placed as in the variable gadget.
Class~$Y$ is a copy of $X$, placed outside the convex hull of $X$ and $B$.
The point $r_1$ is placed inside the convex hull of $B$ but outside the convex hull of $X$.
The point $r_4$ is placed inside the convex hull of $Y$ and $r_2$ and $r_3$ are placed such that the line through them separates the convex hulls of $B$ and $Y$.
Then, either partition drawing on $X$ and $B$ induces a unique partition drawing of $R$ and $Y$, where the drawing on $Y$ is the same as the drawing on $X$.

Placing $Y$ as a copy of $B$ instead of $X$, i.e., with only one point in the convex hull of~$R$, we can also turn this gadget into an inverter gadget.

\paragraph{The splitting gadget.}
The splitting gadget consists of three variable gadgets $X$, $Y$, and~$Z$, and two additional color classes, the \emph{red} color class $R$ and the \emph{blue} color class $B$, see Figure~\ref{fig:split4}.
The truth assignment on $X$ enforces some edges in $R$ and $B$ to be present, which then uniquely determines the partition drawing on the whole gadget. 
Note that the truth assignments on~$Y$ and~$Z$ are enforced as the negated truth assignment on~$X$, so an additional inverter gadget might be needed depending on the required literal.

\begin{figure}[b]
  \centering
  \includegraphics[page=2]{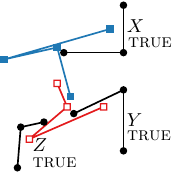}
  \hfill
  \includegraphics[page=3]{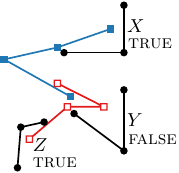}
  \hfill
  \includegraphics[page=4]{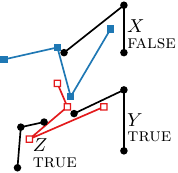}
  \hfill
  \includegraphics[page=5]{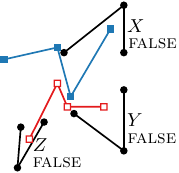}
  \caption{Assignments of the OR-gadget.}
  \label{fig:orassign4}
\end{figure}

\paragraph{The OR-gadget.}
The OR-gadget consists of three variable gadgets $X$, $Y$, and $Z$, and two additional color classes, the \emph{red} color class $R$ and the \emph{blue} color class $B$, see Figure~\ref{fig:orassign4}.
The truth assignments on $X$ and $Y$ enforce some edges in $R$ and $B$ to be present.
It can be seen that the drawing of $Z$ corresponding to the value \true can only be drawn if $X$ or~$Y$ are also drawn corresponding to the value \true.
In some of these cases, $Z$ could also be drawn according to the value \false, but this does not affect the proof as it is still true that the constructed point set admits a partition drawing if and only if the planar \textsc{3SAT} formula~$\varphi$ is satisfiable.

\bibliographystyle{plainurl}
\bibliography{abbrv,partitiontrees}

\end{document}